\documentclass[]{jltp}
\usepackage{graphicx}
\title{
Spectral Properties of High-T$_c$ Cuprates via a Cluster-Perturbation Approach
\footnote{Dedicated to
  Prof. Peter W\"olfle on the occasion of his sixtieth birthday.}}
\author{C. Dahnken,  E. Arrigoni, and W. Hanke}
\address{Institut f\"ur Theoretische Physik, Universit\"at W\"urzburg, \\
Am Hubland,  97074 W\"urzburg, Germany}
\runninghead{C. Dahnken, E. Arrigoni, and W. Hanke}{
  Spectral Properties of High-T$_c$ Cuprates 
  }

\def\v#1{{\bf #1}}
\def\eqref#1{Eq.~(\ref{#1})}
\def\beq{\begin{equation}}
\def\eeq{\end{equation}}

\begin{document}
\maketitle
\begin{abstract}
Angular-resolved photoemission data on half-filled doped cuprate
materials are compared 
with an exact-diagonalization analysis of the 
three-band Hubbard model, which is
extended to the infinite lattice by means of a 
perturbation in the intercluster hopping (cluster perturbation
theory).
A study of the band dispersion and spectral weight
of the insulating cuprate Sr$_2$CuO$_2$Cl$_2$ allows us to fix a
consistent parameter set, which turns out to be appropriate at finite
dopings as well.
In the overdoped regime, 
our results for the spectral weight and for the Fermi surface
give a good description of the experimental data on Bi$_2$Sr$_2$CaCu$_2$O$_{8+\delta}$.
In particular, the Fermi surface is hole-like and centered around
$k=(\pi,\pi)$. 
Finally, 
we introduce a hopping between two layers and
address
the issue of bilayer splitting.

PACS numbers: 74.72.-h, 74.72.Hs, 79.60.-i. 
\end{abstract}
\section{INTRODUCTION}
  Angle-resolved photoemission spectroscopy (ARPES) has
provided significant insight into the single-particle 
properties of high-temperature superconductors (HTSC),
providing  a measure of the $k$-dependent spectral
function below the Fermi surface, which is suitable for a direct
comparison with 
theoretical models.
Recently,  particular  attention has been devoted  to
the spectral properties of
 undoped antiferromagnetic parent compounds of the HTSC, 
mainly Sr$_2$CuO$_2$Cl$_2$ 
due to its ideal experimental characteristics. 
A series of photoemission measurements \cite{duerr2000,laro1997,well1995} 
of these half-filled cuprates provides a detailed view of the 
quasiparticle dispersion of the highest electron-removal state, 
commonly known as the Zhang-Rice-Singlet (ZRS) \cite{zh.ri.88}. 
In a number of previous works,
the wave-vector dependence 
of this band has 
been theoretically analyzed
on basis of exact-diagonalization (ED) and quantum monte-carlo (QMC) 
studies of Hubbard \cite{preuss1995}, t-J \cite{eder1997} and three-band models \cite{dopf90,dopf92-1,dopf92-2}. 
On the other hand, 
it is clear that an appropriate inclusion of the most relevant 
orbitals 
(Cu3d$_{x^2-y2}$, O2p$_{x}$ and O2p$_{y}$) of the CuO$_2$ 
unit cell, i. e., based on the three-band Hubbard model
should provide a more complete description of the system.
The trouble is that an exact-diagonalization analysis of this model 
has to be necessarily
restricted to a
rather limited number of unit cells,
due to the rapid
increase of the size of the Hilbert space. Even QMC simulations 
of larger systems were still severly restricted, 
providing informations
on only a few k-points in the Brillouin zone.

To remedy this problem, we  follow here
the idea \cite{senechal2000} 
of extending
 the results of the exact
diagonalization of a small cluster 
to the whole
lattice within a lowest-order perturbation in the intercluster hopping.
 This technique has been shown to produce quite good results for 
the Hubbard~\cite{senechal2000} model
and has been successfully applied 
to the study of the stripe phase in the cuprate materials
\cite{zacher2000,zacher2001}. 
It has the advantage
to treat accurately short-range correlations, which are believed to be the
most important ones in these systems~\cite{fulde}.
Furthermore, 
at the same time, it accesses all $k$ points 
of the Brillouin zone, while in ordinary ED only few $k$ points are
available.
This last property is, of course, crucial if one wants to analyze a
dispersion  or determine a Fermi surface with sufficient details.

As a matter of fact, recent advances in experimental ARPES 
techniques allowed for 
a direct mapping of the Fermi surface, which has been experimentally analyzed
in detail in overdoped Bi$_2$Sr$_2$CaCu$_2$O$_{8+\delta}$ (Bi2212) 
\cite{fret2000,grom2000,feng2001,meso2001,gold99,bogd2001,mars1996}. 
Unfortunately, there seems to be disagreement between  results
obtained  at different
photon energies.
Measurements taken at a photon energy of 22eV
typically show a Fermi surface closed around ${\bf k}=(\pi,\pi)$ 
("hole-like"), whereas at 33eV photon energy the 
same materials show a shape closed around
${\bf k}=(0,0)$ ("electron-like").
To understand this issue, we extend our analysis
to an overdoped system with 25 \% doping, where we keep the same
parameters obtained at
half filling.
Our results give 
a quasiparticle dispersion which is
compatible with the experiments  on the overdoped cuprates,
exhibiting a hole-like Fermi surface closed around $\v k=(\pi,\pi)$.

In the final part of the paper,
we present results obtained by coupling two layers 
with an interlayer hopping 
$t_{\perp}=0.1 t_{pd}$, and study its effects on the
Fermi-surface splitting, comparing with experimental results.


\section{MODEL AND TECHNIQUE}
  Common 
property of all HTSC and of their parent compounds 
is the presence of stacked
CuO$_2$ layers. The general agreement is
that the relevant
excitations of these materials  take place within these layers.
A generic model considers
three relevant 
orbitals, namely,  Cu$_{d_{x^2-y^2}}$, O$_{p_x}$ and O$_{p_y}$ orbitals\cite{bren1995,fulde,shen.dessau1995}. 
Restricting to the on-site Coulomb interactions within the $d$
copper orbitals only,
this gives the three-band Hubbard Hamiltonian \cite{emery1987_88}
\begin{eqnarray}
  \label{eq:3bham}
  H_{3b}&=&  -t_{pd} \sum_{\left< ij \right>,\sigma} 
  \alpha_{ij} \left( d^\dagger_{i,\sigma} p_{j,\sigma}
    +p^\dagger_{j,\sigma} d_{i,\sigma}  \right) \cr
  &&-t_{pp} \sum_{\left< jj' \right>,\sigma} 
  \alpha'_{jj'} \left( p^\dagger_{j,\sigma} p_{j',\sigma}
    +p^\dagger_{j',\sigma} p_{j,\sigma}  \right)\cr
  && +\frac{U_d}{2}\sum_{i,\sigma,\sigma'} 
  d^\dagger_{i,\sigma}d_{i,\sigma}d^\dagger_{i,\sigma'}d_{i,\sigma'} \cr 
  &&+\Delta \sum_{j,\sigma} 
    p^\dagger_{j,\sigma}p_{j,\sigma},
\end{eqnarray}
where the operators $d^\dagger_{i\sigma}$ and $p^\dagger_{j,\sigma}$ 
create  holes in the Cu 3d and in the O 2p orbitals, respectively, and
$\alpha_{ij}$ and $\alpha'_{jj'}$ give the usual
orbital phase factors. $\left<...\right>$ denotes summation over 
nearest neighbours.

Exact diagonalizations of  $H_{3b}$ are usually 
limited to $4$ unit cells,
allowing for just $3$ inequivalent 
wave vectors, which are not enough 
to address questions of 
quasiparticle dispersion or of Fermi surface. 
Nevertheless, 
since the important 
correlations  of these strongly-correlated materials should be relatively
short range (a celebrated example is the Zhang-Rice singlet~\cite{zh.ri.88})
one can still envisage to capture these short-range correlations by
exact diagonalization of a small cluster, and then continue the cluster
properties to the infinite lattice within a perturbation in the
inter-cluster hopping elements.
This idea has been suggested by S{\'e}n{\'e}chal and
coworkers\cite{senechal2000}  and has 
been termed ``Cluster-Perturbation Theory''(CPT).
A closely related 
 method 
has been previously suggested 
 to treat the coupling
between one-dimensional Luttinger liquids, and to address the problem of
the crossover from one to higher
dimensions~\cite{cross.97,bo.bo.95,arri.99.c}. 

Central piece of this method 
is the Green function of the cluster $G_{ij\sigma}(z)$, which is
obtained from Lanczos diagonalization. At the lowest order
in the CPT, the full 
cluster 
Green's function is non-diagonal in the momenta, as the partition into 
clusters produces a smaller ``superlattice'' Brillouin zone (BZ).
More specifically, one has
\beq
{\cal G}^{CPT}_{{\v q+ \v G_1},{\v q+ \v  G_2},\sigma}(z)
=
\frac{1}{N_0} \sum_{\v r_1,\v r_2}^{cluster}
 e^{i \v q \cdot \left({\v r_1}-{\v r_2} \right)} 
 e^{i \v G_1 \cdot {\v r_1}  -i \v G_2 \cdot \v r_2}
{\cal G}_{\v r_1,\v r_2,\sigma}({\v q},z) \;,
\label{gcpt}
\eeq
where $\v G_i$ are vectors of the reciprocal superlattice, $\v r_i$ are
cluster sites,  $\v q$ is restricted within the superlattice BZ, and $N_0$ 
is the number of unit cells  of the cluster.
Here, the cluster Green's Function in the superlattice representation 
${\cal G}_{\v r_1,\v r_2,\sigma}({\v q},z)$ can be written as
\beq
{\cal G}_{\v r_1,\v r_2,\sigma}({\v q},z) = \left[{\v G}_{\sigma}(z)^{-1} -
  {\v V} ({\v q})\right]^{-1}_{\v r_1,\v r_2} \;,
\label{gcpt1}
\eeq
where both the cluster Green's function
 ${\v G}_{\sigma}(z)$ as well as the 
Fourier-transformed
intercluster hopping 
$\bf{ V(\v q)}$ are
 $N_0 \times N_0$
matrices in the cluster-site indices.
$V(\v q)$ is the only term connecting different clusters and, thus,
displaying a $\v q$-dependence.
More specifically, let $T_{\v r_1,\v r_2}(\v \Delta)$ be
the  amplitude of hopping of a particle from  a site $\v r_1$ in  cluster
$\v R$ and a site $\v r_2$ in cluster $\v R + \v \Delta$ 
(which,
 for a translation-invariant system,
 only depends on 
the relative position, 
$\v \Delta$,
of the two clusters)
then $\v V(\v q)$ is 
given by
\[
  V(\v q)_{\v r_1,\v r_2} = 
\sum_{\v \Delta} e^{-i \v q \cdot \v \Delta} \ 
T_{\v r_1,\v r_2}(\v \Delta) \;.
\]
This method obviously 
 allows for a perturbative treatment of {\it intracluster}
hopping terms as well.
These can be included via the intercluster part
 $T_{\v r_1,\v r_2}(\v \Delta=\v 0)$.
This gives the possibility of
diagonalizing the cluster with {\it periodic}
boundary conditions (BC), which is, clearly, more convenient
numerically,
but which contains unphysical hopping terms within the cluster.
These additional terms, which are not present in the original problem,
are then  removed perturbatively via a 
$T_{\v r_1,\v r_2}(\v \Delta=\v 0)$ term with the opposite sign.

The non-diagonal terms $\v G_1\not= \v G_2$ in \eqref{gcpt} 
 are produced by the different treatment of the intra- and
 intercluster hopping 
terms. 
Within the CPT, these terms are neglected, and one takes only the
$\v G_1=\v G_2$ part of the Green's function to evaluate the spectral
properties.
Of course, in our multiband Hubbard model, the Green's function is additionally a
$3\times 3$ matrix in the orbitals of a single cell.
The spectral function plotted in figures \ref{fig:hfdisp} and \ref{fig:disp0.75} 
is given by the sum of the diagonal elements of the Green's-function matrix.

Notice,  that the lowest-order CPT becomes exact  in
the trivial case of
vanishing  intercluster hopping and for non-interacting systems.
This is due to the fact that corrections to \eqref{gcpt} with
\eqref{gcpt1} are given by higher-order cumulants, which are vanishing
when Wick's theorem holds, i. e. for non-interacting electrons.
This fact makes the CPT an appealing interpolation between the strong- 
and the weak-coupling limits.

Our cluster consists of $N_0=2\times2$ unit cells of the $3$-band
model, for a total of $12$ lattice sites. For a single doped 
hole, the calculation of the Green's Function requires 156
diagonalizations of $H_{3b}$ with at most 48400 basis states.

In principle,
if we require that the hole density of the cluster fixes the doping of the
system, the lowest doping we that can treat (beyond half filling) 
is $\delta=\frac14$. Even so,
we have to average the Green's function between the two situations
where the additional particle in the cluster has spin up or down.

As a matter of fact, one can also treat any doping  between $\delta=0$ and
$\delta=25\%$,
 by considering 
the cluster as being in a grand-canonical ensemble with fixed
chemical potential $\mu$  between hole numbers
$n_h=4$ and $n_h=4+1$. For finite temperature $T$, one can always fine-tune
$\mu$ so that the {\it average} $n_h$ is the desired one
$n_h=4(1+\delta)$. In the $T \to 0$ limit, this is obtained by using
as a cluster Green's function 
to insert in \eqref{gcpt1}
the weighted sum of the two Green's functions at 
$n_h=4$ and $n_h=5$
($G_{n=4}$ and $G_{n=5}$, respectively), i. e.,
$G_{<n_h>=4+x} = x \ G_{n_h=5} + (1-x) \ G_{n_h=4}\quad \quad$ ( with 
$0\leq x\leq1$),
 at the appropriate chemical
potential. 
This is, physically, what one expects to occur in a cluster embedded in 
a larger system with $\delta=\frac{x}{4}$, namely, the particle number 
of the cluster $n$ will mainly
fluctuate between $n=4$ and $n=5$.


\section{RESULTS}
\begin{figure}[tbp]
  \begin{center}
    \includegraphics[scale=0.42]{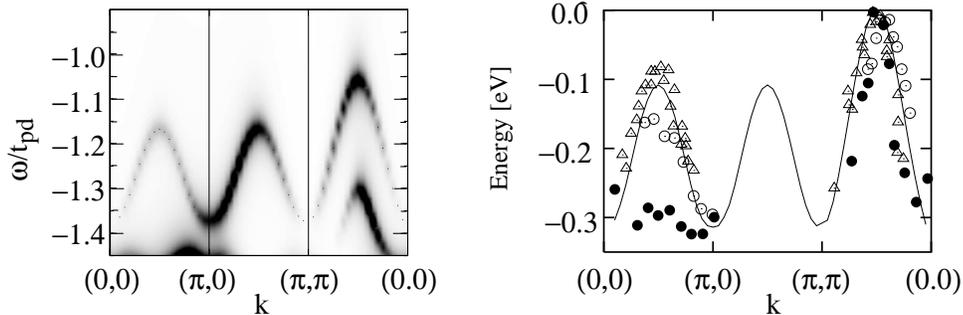}    
    \caption{Grayscale plot of the 
      spectral density of the half filled three-band Hubbard
      model, 
      \eqref{eq:3bham}, 
      with
      $t_{pp}=0.4$, $U_d=6$ and $\Delta=4$ (in units of $t_{pd}=1.3eV$),
      as
      obtained from CPT (left).  The points 
      mark the peak positions.
      The right panel shows the band dispersion obtained from
      different experiments (circles \cite{duerr2000}, filled circles
      \cite{mars1996} 
      and triangles \cite{laro1997}).
}
\label{fig:hfdisp}
  \end{center}
\end{figure}

  First, we will discuss the dispersion of the highest electron 
removal states on the basis of the most recent experimental 
studies of the half-filled cuprate Sr$_2$CuO$_2$Cl$_2$.
Up to few exceptions \cite{well1995}, these studies provide a 
consistent picture of the quasiparticle dispersion of the 
HTSC parent compounds \cite{duerr2000,laro1997},
showing two parabolas centered at $k=(\pi/2,0)$
and $k=(\pi/2,\pi/2)$. A characteristic energy difference  
$\Delta_a \approx 100-150meV$ between the maxima of the parabolas
is observed. 
The total width W of the first band  is about $300-400meV$ 
(compare figure \ref{fig:hfdisp} for experimental data).
In accordance with 
earlier numerical studies \cite{dopf90,dopf92-1,dopf92-2,bren1995}
of the three-band model,
we find 
that the parameter set
$t_{pp} = 0.4 t_{pd}$ ,
$U=6 t_{pd}$ and  $\Delta=4t_{pd}$ gives the best agreement with the
experimental band structure.
 With the  value of the 
copper-copper hopping matrix element\cite{bren1995} taken from the
literature,
 $t_{pd}=1.3eV$, our 
calculations yield $\Delta_a=150meV$ and $W=400meV$. For the charge transfer 
gap we get $\Delta_{ct}=1.8eV$ in good agreement with 
the experimental value of
 $1.5-2eV$ \cite{uchida1991,cooper1993}. 
All these quantities are 
well within the 
margins of the experimental constraints.
Therefore, these parameters should give a rather
 good description of the materials.
%
%
Figure \ref{fig:hfdisp} shows a density plot of the single-particle 
spectral function $A(k,\omega)$ along the high-symmetry directions of the 
Brillouin zone (left panel).
On the right panel we show corresponding ARPES data for 
comparison\cite{well1995,duerr2000,laro1997}.
The dispersion of the band is reproduced quite well by our
calculation.
On the other hand, the
spectral weight is too strong in the region $(\pi,0)$ to $(\pi,\pi)$,
and too weak in the region between $(0,0)$ to $(\pi,0)$.
This apparent disagreement is discussed in more detail below.
 Note that the scale of the binding energy is not absolute, since we
 are dealing with insulators.

We now comment on previous analyses of the experimental band dispersion 
displayed in Fig.~\ref{fig:hfdisp} which have been based on the t-J model.
This model is believed to be an effective low-energy version of the 
three-band Hubbard model\cite{hybe1990}. The t-J model is more
appropriate for numerical analysis, since his Hilbert space is smaller
for a given number of unit cells.
However, our results obtained by a CPT analysis of the three-band Hubbard model
clearly differs from what one generally obtains from the t-J model.
The latter predicts, in contrast to experiments, 
a flat band close to the Fermi energy at $(\pi,0)$ and $(\pi,0)-(0,\pi)$ \cite{tohy2000}.
A solution to this shortcoming of the t-J model 
has been suggested, by introducing longer-range hopping
and interaction terms.
In fact, a mapping of the three-band Hubbard model
into a generalized t-J model has been recently  
proposed \cite{erol1998}, which is consistent with results of 
the self-consistent Born approximation calculations  and
experimental data \cite{lema1998,tohy2000,dama2001}.  In particular,
these latter results 
exhibit a dispersion in excellent agreement with our calculations of 
 Fig.~\ref{fig:hfdisp}. 

From Fig.~\ref{fig:hfdisp}, one can see that our calculation
underestimates the spectral weight along 
$(0,0)-(\pi,0)$, which is usually  strong in ARPES.
One reason could be related to the
small cluster in which the self-energy is evaluated, which takes only partially
into account the electron correlations.
Moreover, it was established numerically \cite{senechal2000} that quasiparticles
contained in the cluster are reproduced correctly, but they loose 
coherence over the cell boundary. 
Alternatively, the different spectral weight could be caused by 
intrinsic limitation of the model. For example, neglected orbitals \cite{ande1995} -- 
although distant in energy --
could possible give significant contributions to the spectral weight.
%
%
%
%
\begin{figure}[tbp]
  \begin{center}
      \includegraphics[scale=0.4]{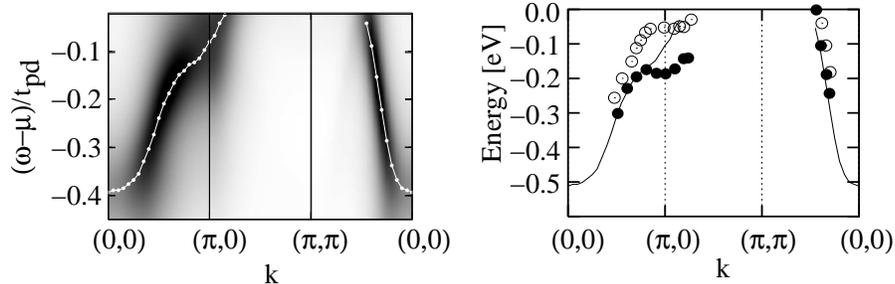} 
    \caption{Spectral density of the overdoped 
      ($\delta=0.25$) three-band Hubbard model as obtained by CPT
      (left panel).
      Parameters are as in Fig.~\ref{fig:hfdisp}.
      The white dots  mark the 
      peak positions. Corresponding experimental results from 
      Ref. \onlinecite{mars1996}
      are included for comparison (right panel). Empty circles denote 
      the overdoped, filled circles the underdoped case. The solid line 
      marks the CPT result.}
    \label{fig:disp0.75}
  \end{center}
\end{figure}
\begin{figure}[tbp]
  \begin{center}
    \includegraphics[width=15cm]{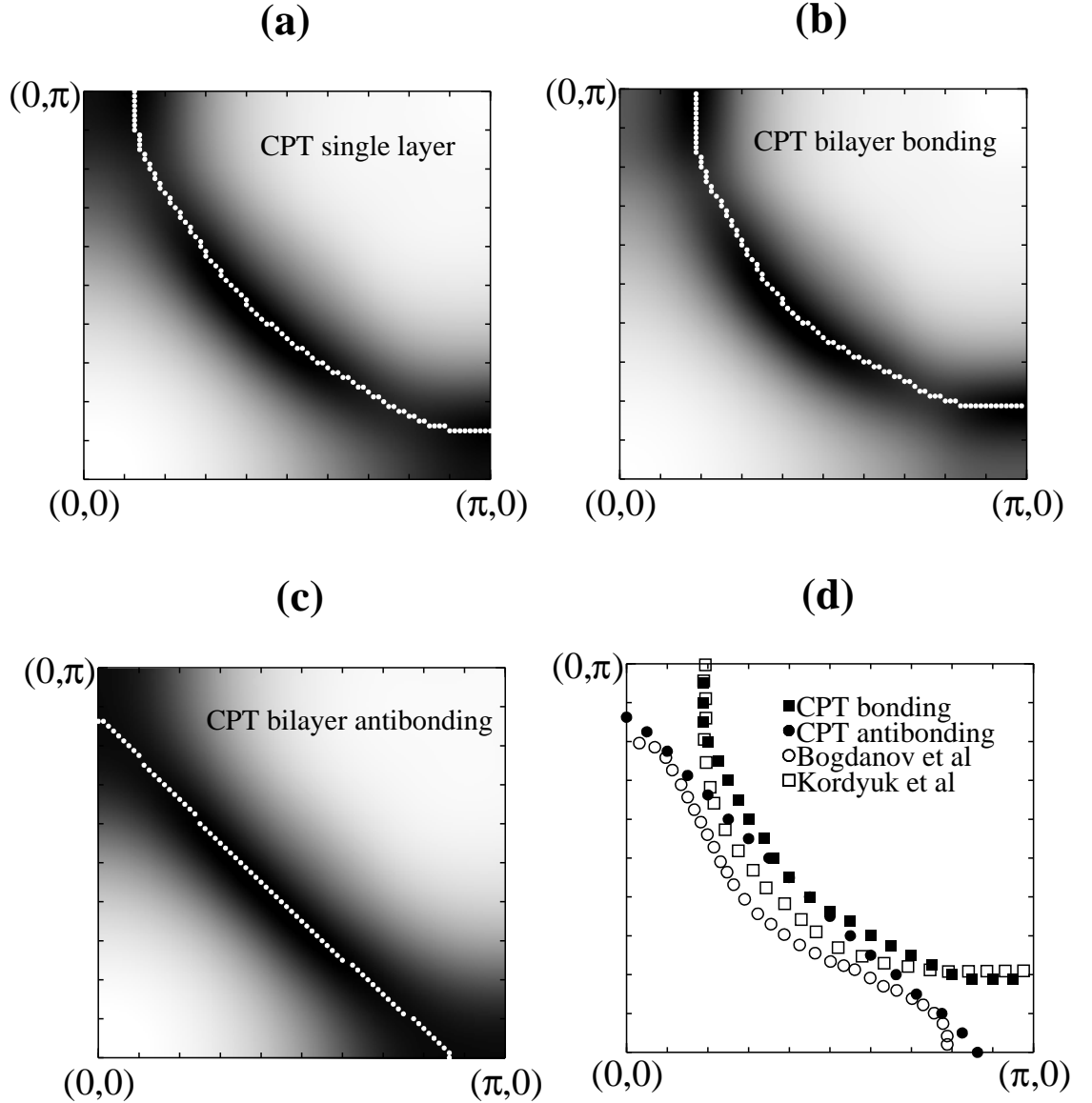} 
    \caption{Spectral weight at the Fermi energy giving 
      an estimate of the position of the Fermi surface.
      Results are displayed for a single CuO$_2$ layer (a)
      and for the bonding (b) and antibonding (c) band
      of a bilayer with interlayer hopping $t_{\perp}=0.1t_{pd}$.
      The white dots indicate the maxima of the spectral weight on a
      given branch.
      Panel (d) shows the comparison with experimental results on 
      overdoped (Bi,Pb)-2212  at 22eV photon energy \cite{kordyuk2001}(empty squares) 
      and and slightly overdoped (Bi,Pb)-2212 at 55eV photon energy \cite{bogd2001}(empty circles).
      Filled squares and circles denote the results of the CPT calculation
      for the bonding and antibonding branch, respectively.
      }
    \label{fig:fs0.75}
  \end{center}
\end{figure}

We now turn to the doped systems. Here, we assume, as a first guess,
 that the parameters
are not significantly dependent on the doping and on the particular material, 
and we keep the same ones for the doped region.
The minimum doping that can be achieved with a 12-site cluster by 
adding an additional electron is $\delta=0.25$. This undoubtedly high
doping is nevertheless well suited to be compared to the detailed 
experimental results on dispersion \cite{mars1996} and Fermi surface 
\cite{fret2000,grom2000,feng2001,meso2001,gold99,bogd2001}
of the overdoped cuprates. The results of our calculation are plotted 
in Fig.~\ref{fig:disp0.75}. The  quasiparticle dispersion from 
$k=(0,0)$ to $k=(\pi,\pi)$  exhibits a steep ascent beginning 
at a binding energy of $\sim -0.5eV$ and crossing the Fermi energy 
shortly before $k=(\pi/2,\pi/2)$. Starting from $k=(0,0)$ and going
along (1,0), the dispersion reaches a saddle point at $\sim k=(\pi,0)$ well 
below the Fermi energy at $\sim -100meV$. 
A similar saddle-point dispersion is
observed in the experimental
data for Bi$_2$Sr$_2$CaCu$_2$O$_{8+\delta}$ \cite{mars1996}, 
shown for comparison in the right panel of Fig.~\ref{fig:disp0.75}.
The good qualitative agreement of our numerical results with the
experimental ones suggests that our approach of taking the 
same parameter values in the undoped and in the overdoped systems
is quite reasonable. This was also found earlier by intensive 
QMC simulation \cite{dopf92-1}.

The magnitude of the spectral weight at the Fermi energy is  
displayed as a grayscale plot in Fig.~\ref{fig:fs0.75} 
for the whole Brillouin zone.
The darkest regions, thus, give an estimate of the location of the Fermi
surface.
The  data clearly show a hole-like barrel closed
around $k=(\pi,\pi)$, as usually observed in ARPES experiments on 
the overdoped cuprates \cite{fret2000,meso2001,kordyuk2001,legner2000}, 
mostly at a photon energy of $E_\nu\approx 22eV$.
On the other hand, experiments carried out 
at a higher energy \cite{grom2000,bogd2001,chuang1999}
$E_\nu=33eV$ disagree with these results, and report an electron-like
Fermi surface. This is quite puzzling, and there is not yet a full
consensus about this issue.
Some recent papers suggest that the observation of 
two different Fermi surfaces is  actually due to the
splitting from the coupling between the bilayers \cite{bogd2001,feng2001}.
Nevertheless,  our data  suggest that, if the bilayer splitting is not relevant, 
the Fermi Surface should be hole-like. 

The coupling between two layers  can be treated 
without  further difficulties within our method. Again, we diagonalize a
$2\times2$ unit cell exactly and include an interlayer hopping term  
$t_{\perp}$ within the CPT treatment.
Since the effective hopping $t_{\perp}$ under consideration is
relatively small and  
CPT amounts to a perturbation 
in the intercluster hopping, 
we believe that this is a good procedure. 

The interplane hopping has been analized in 
studies based on LDA calculations and has been mapped 
on effective multi-orbital models by integrating out 
high-energy degrees of freedom \cite{ande1995}. This
approach provides results which are  consistent with the
experimentally observed bilayer splitting \cite{feng2001}. 
In our calculation, we
  adopt this idea by treating the orbital phase factors of the 
various interplane hoppings accordingly.
We take a value of $t_{\perp} \approx 0.1 t_{pd}$, which is 
sufficient to produce an observable splitting of the Fermi surface.

Figure \ref{fig:fs0.75} (right panel) displays our results 
obtained for the bilayer.
Different experimental results are shown in the last panel for comparison.
Two branches  can be distinguished, one being almost 
a square closed around $k=(\pi,\pi)$,
The other one is  closing around 
${k=(0,0)}$, forming an electron-like 
Fermi surface as observed in Ref.~\onlinecite{bogd2001}. 
Similar results have also been found 
by first-principle calculations of ARPES 
in Bi2212 \cite{bansil1999}.
These results suggests, that a possible reason for 
the crucial differences in the experimental observations
might be found in interlayer processes. 
An explanation of the apparent 
dependence of the Fermi surface on the
photon energy
would require the consideration of 
matrix element effects which have not been addressed 
within this paper, but are treated in detail elsewhere \cite{da.ed.01u}.
%
%


\section{CONCLUSION}
This paper reports a systematic study of single-particle spectral
properties of the three-band Hubbard model in the half-filled and 
in the doped region.
In order to obtain sufficiently accurate results and 
to sort out details of the band dispersion we adopt a method
(so-called CPT) 
combining
exact diagonalization of  small clusters with a perturbation in the
intercluster hopping. 
This analysis allows us to determine a parameter set, which appropriately
describes both the insulating as well as the overdoped region and gives a
good agreement with ARPES data.
The Fermi surface obtained at high dopings is hole like, although  
an interlayer hopping of about $t_{\perp}\approx 0.1t_{pd}$
produces a splitting into an electron and a hole-like Fermi surface
, in agreement with ARPES experiments. 
\section*{ACKNOWLEDGMENTS}
We thank R. Eder for useful discussions.
This work was supported by the projects DFG HA1537/17-1 and HA1537/20-1, 
KONWIHR OOPCV and BMBF 05SB8WWA1.
Support by computational facilities of HLRS Stuttgart and LRZ Munich
is acknowledged.


\bibliographystyle{myprsty-jltp}
\bibliography{biblio,articles,preprints,mypublications,books}

\end{document}